\documentclass[rnote]{aa}
\usepackage{graphicx}

\newcommand{\pri}    {${\rlap.}^{\prime \prime}$}
\newcommand{\rl}     {${\rlap.}^{s}$}

\newcommand{\ltsima} {$\; \buildrel < \over \sim \;$}
\newcommand{\simlt}  {\lower.5ex\hbox{\ltsima}}            
\newcommand{\gtsima} {$\; \buildrel > \over \sim \;$}
\newcommand{\simgt}  {\lower.5ex\hbox{\gtsima}}            

\begin{document}

\title{A catalogue of ULX coincidences with FIRST radio sources}

\author{
J.~R. S\'anchez-Sutil\inst{1}
\and A.~J. Mu\~noz-Arjonilla\inst{1}
\and J. Mart\'{\i}\inst{2,1}
\and J.~L. Garrido\inst{2,1}
\and D. P\'erez-Ram\'{\i}rez\inst{2,1}
\and P. Luque-Escamilla\inst{3,1}
}

\offprints{J. Mart\'{\i}}

\institute{
Grupo de Investigaci\'on FQM-322, 
Universidad de Ja\'en, Campus Las Lagunillas s/n, Edif. A3, 23071 Ja\'en, Spain \\
\and
Departamento de F\'{\i}sica, EPS,  
Universidad de Ja\'en, Campus Las Lagunillas s/n, Edif. A3, 23071 Ja\'en, Spain \\
\email{jmarti@ujaen.es, jlg@ujaen.es, dperez@ujaen.es} 
\and Dpto. de Ing. Mec\'anica y Minera, EPS,
Universidad de Ja\'en, Campus Las Lagunillas s/n, Edif. A3, 23071 Ja\'en, Spain \\
\email{peter@ujaen.es}
}

\date{Received / Accepted}

\titlerunning{A catalogue of ULX coincidences with FIRST radio sources}

\abstract
{}
{We search for ultra luminous X-ray source (ULXs) radio counterparts
located in nearby galaxies in order to constrain their physical
nature.}
{Our work is based on a systematic cross-identification of the most
recent and extensive available ULX catalogues and archival radio
data.}
{A catalogue of 70 positional coincidences is reported. Most of them
are located within the galaxy nucleus. Among them, we find 11 new
cases of non-nuclear ULX sources with possibly associated radio
emission.}
{}

\maketitle

\section{Introduction} 

Ultra luminous X-ray sources (ULXs) are common phenomena nowadays.
Detected in several galaxies as extranuclear X-ray sources, their
luminosities are often in excess of $\sim 10^{39}$ erg s$^{-1}$ in the
soft X-ray bands as observed by several satellite observatories
($ROSAT$, $XMM$, etc.). Their existence was observationally stablished
nearly two decades ago (Fabbiano 1989).  Also known as
intermediate-luminosity X-ray objects (IXOs), their physical nature
still remains controversial. Unless beaming effects were at work,
their intrinsic luminosities are apparently in between those of normal
X-ray binaries (e.g. $\sim 10^{38}$ erg s$^{-1}$) and X-ray emitting
active galaxies (e.g. $\sim 10^{43}$ erg s$^{-1}$ in the M87 example).

Several hypothesis have been proposed to explain such extraordinary
objects, from accreting intermediate mass black holes (IMBH) having
$\sim 10^2$-$10^4$ $M_{\odot}$ to microquasar X-ray binaries with
strong beaming effects (i.e. microblazars). In particular, the IMBH
interpretation has been recently favoured, based on X-ray variability
studies and spectral fits in the M74 ULX, and supported by the
estimated orbital period in M82 X-1, which implies black holes of
thousands of solar masses (Liu et al. 2005; Kaaret et al. 2006). The
reader is referred to \cite{f2004}, \cite{mc04}, \cite{s2005}, and
\cite{f2006} for some recent reviews.

Progress towards a solution to the ULX problem has been possible in a
few cases for which a candidate radio counterpart was detected. At
radio wavelengths, extinction by the interstellar medium is not an
issue in contrast to other domains of the electromagnetic spectrum.
Moreover, modern radio interferometers provide excellent angular
resolution and sensitivity when searching for faint radio counterparts
and studying their spectral index and variability properties.  The
case of the ULX in NGC 5408 reported by \cite{k03} is a prototype
example of this approach followed by other radio works (K\"ording et
al. 2005; Miller et al. 2005).

Unfortunately, the cases for which radio counterparts are available
are not significant enough to establish statistically robust
conclusions. Therefore, efforts to increase the number of radio
detections could potentially lead to further evidence.

The total number of ULX objects has been growing extensively in past
years. In particular, two extensive catalogues of ULXs have recently
become available from \cite{lm2005} and \cite{lb2005}, containing
several hundreds of these objects. This fact, together with the
availability of sensitive radio surveys at high galactic latitudes,
makes it feasible and worthwhile to attempt a systematic
cross-identification search for new cases of ULX with associated radio
emission. The Faint Images of the Radio Sky (FIRST) survey carried out
with the Very Large Array (VLA) at 20 cm becomes specially suitable
for such a purpose given both its good sensitivity (rms $\sim$ 0.15
mJy) and angular resolution ($\sim 5^{\prime\prime}$), together with
its associated catalogue compiled by \cite{w1997}.

In this context, our goal in the present paper is to present the
results of a search based on a systematic cross-identification of the
whole $\sim 8 \times 10^5$ FIRST radio sources in the most recent
catalogues of the ULX sources quoted above. This approach is similar
to the one carried out by Miller \& Mushotzky that covered nearby
galaxies observed by Chandra and XMM as cited in \cite{mus04}.  In our
case, several new positional coincidences of FIRST radio sources with
ULXs have been found.  They did not appear to be previously recognized
as such in the literature according to the SIMBAD database.  The
following sections are devoted to describing the cross-identification
method and presenting the results. Both radio maps and individual
discussion of coincidence cases of special interest are also included.

\section{Cross-identification method and results}

\begin{table}
\caption[]{\label{catalogos} List of ULX catalogues cross-identified with FIRST}
\begin{tabular}{lccc}
\hline
\hline
                  &      Number of             &   Number of         &   Desig-        \\
catalogue           &      Galaxies              &      ULX            &   nation      \\
                  &                            &                     &                 \\
\hline
\cite{cp2002}     &       54                   &      87             &       CP        \\
\cite{lm2005}     &       85                   &     229             &       LM        \\
\cite{lb2005}     &      313                   &     562             &       LB        \\
\hline
\hline
\end{tabular}
\end{table}

The ULX catalogues used in this work are listed in Table
\ref{catalogos}.  Among them, the most extensive one is that by
\cite{lb2005}, which became the main reference for our search as it
includes most entries. Our search is naturally restricted to the
galaxies in the FIRST radio survey, which implies, for instance, that
only 46\% of the \cite{lb2005} ULXs are covered.

Our cross-identification was based on the positional coincidence of
the FIRST and ULX entries within less than $10^{\prime\prime}$.  This
value comes from the typical positional uncertainty of the ROSAT data
on the X-ray catalogues. The FIRST positional accuracy is around or
better than $1^{\prime\prime}$, i.e., a value that is negligible
compared to that of ROSAT. The cross-identification was carried out in
practice by means of a fortran code that reads and compares the FIRST
and ULXs catalogues previously downloaded from electronic public
archives.

A total of 70 matches resulted from the cross-identification, which
represents 27\% of the ULX sources in the \cite{lb2005} reference
catalogue covered by FIRST. We list all of them in Table \ref{coin2}
in electronic format, where the remarks column indicates coincidences
with the galaxy nucleus, extended radio emission, and previous
identifications in the SIMBAD database. At the FIRST angular
resolution, the expected appearance of a ULX radio counterpart is that
of a compact source for most plausible ULX physical scenarios (IMBH,
microblazar, etc.).  Nevertheless, several extended nebulae have been
recently reported around some ULX sources (Pakull et al. 2003; Kaaret
et al. 2004; Miller et al. 2005), which would be better explained in a
scenario where an IMBH isotropically excites a bubble nebulae around
the ULX. Hence, coincidences with extended radio sources were finally
kept (17\% of matches).

Most entries in Table \ref{coin2} (80\%) overlap with the galaxy
nucleus and, therefore, they are probably not true ULX sources but
simply X-ray emission from the galaxy's central black hole.  Among
non-nuclear matches, only a few that are identified in the SIMBAD
database are up to date.  The remaining 11 coincidences appear as the
interesting sources to be reported here as they represent new radio
counterpart candidates of non-nuclear ULX sources (see Table
\ref{coin}).  The nature of these objects is tentatively suggested
based on their location and morphology.  Thus, a QSO candidate is
considered for compact or double radio sources at a peripheral
galactic location. On the other hand, coincidences of extended radio
emission within knotty optical features or spiral arms are taken as
likely HII regions.

\begin{figure*}
   \centering
\resizebox{\hsize}{!}{\includegraphics{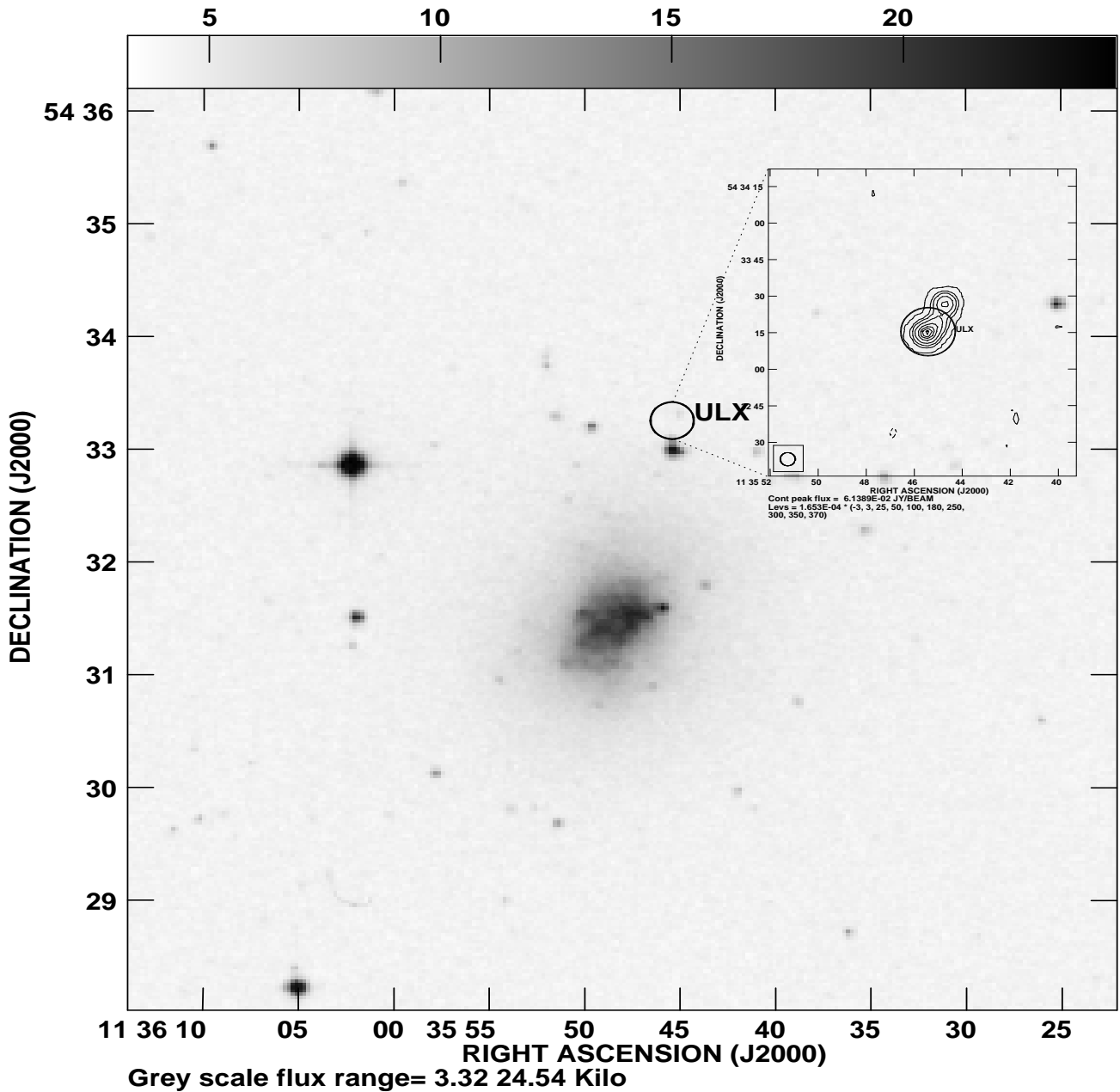}}
      \caption{Radio counterpart candidate of X1 in the field of NGC
3738.  The radio image in the inside panel has been contoured in
units of the FIRST rms noise, typically 0.15 mJy beam$^{-1}$. The
circle in the panel's lower left corner corresponds to the size of the
synthesized beam, about $5^{\prime\prime}$. These values also apply
to the rest of the figures.  }
      \label{ngc3738}
   \end{figure*}

\begin{figure*}
   \centering
\resizebox{\hsize}{!}{\includegraphics[angle=0]{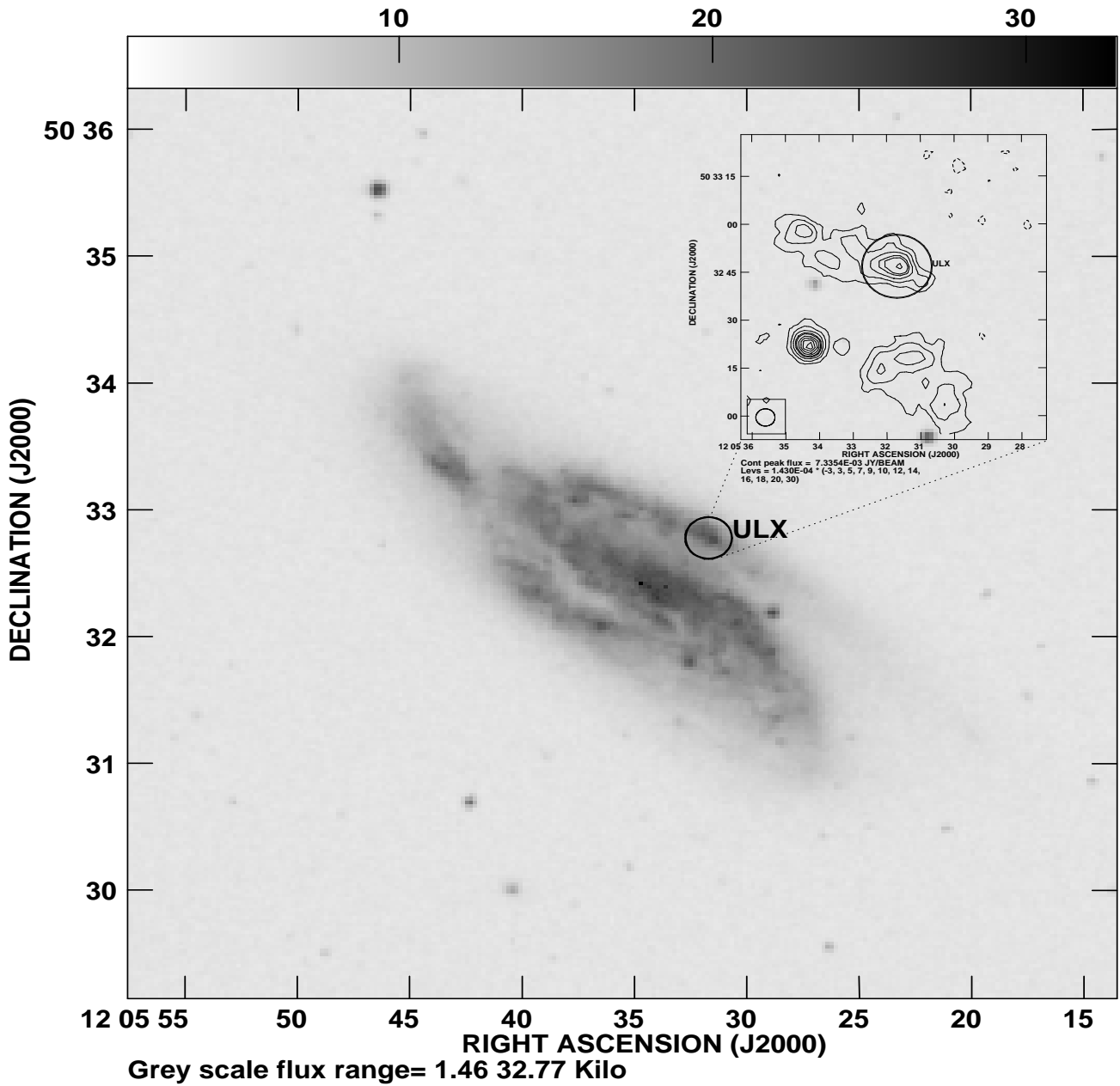}}
      \caption{Coincidence of X1 in the field of NGC 4088 with a
conspicuous peak of radio emission in the galaxy spiral arm.}
      \label{ngc4088}
   \end{figure*}

\begin{figure*}
   \centering
\resizebox{\hsize}{!}{\includegraphics[angle=0]{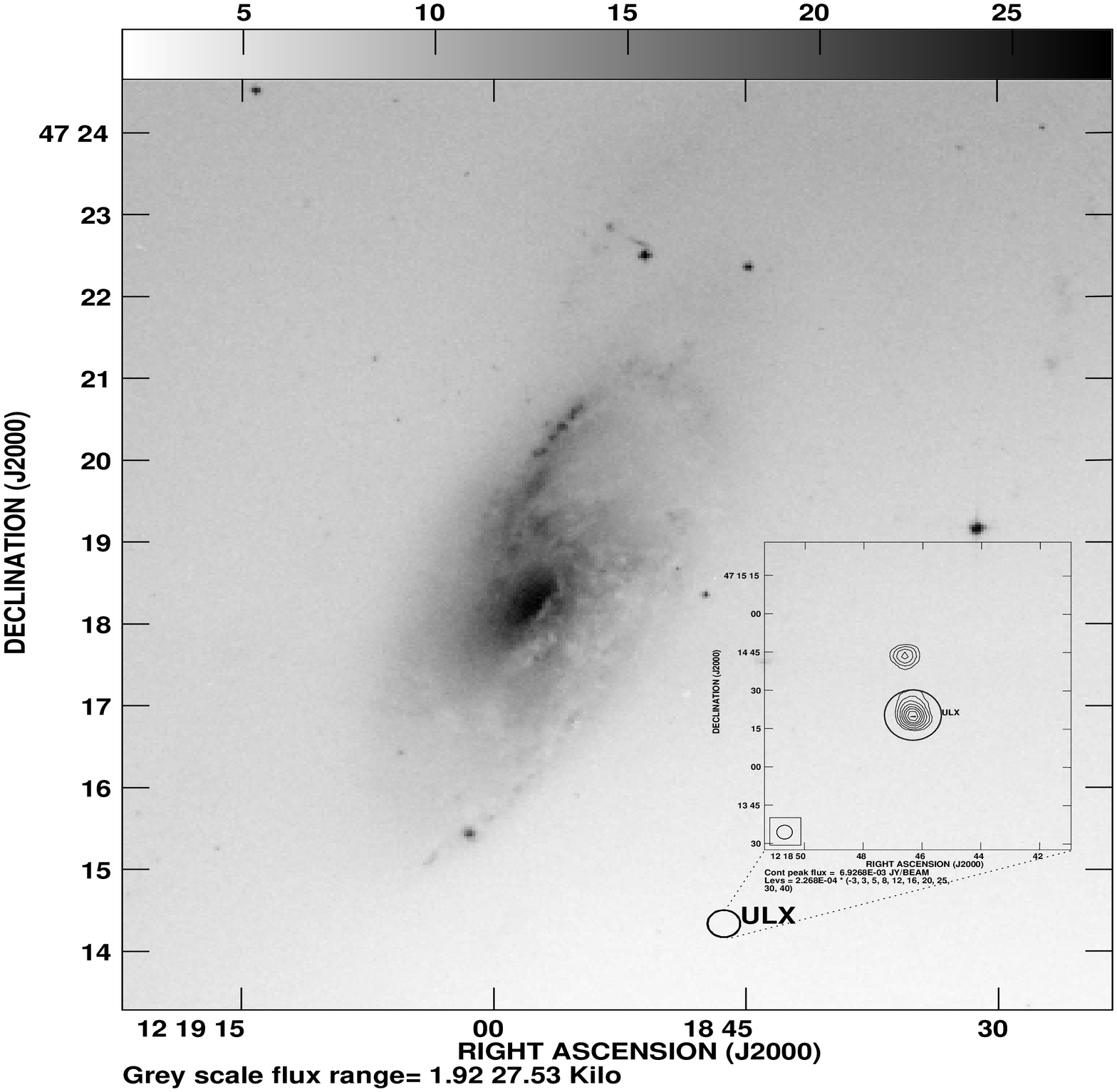}}
      \caption{Radio counterpart candidate of the X6 source in the
      field of NGC 4258.}
      \label{ngc4258}
   \end{figure*}

\begin{figure*}
   \centering
\resizebox{\hsize}{!}{\includegraphics[angle=0]{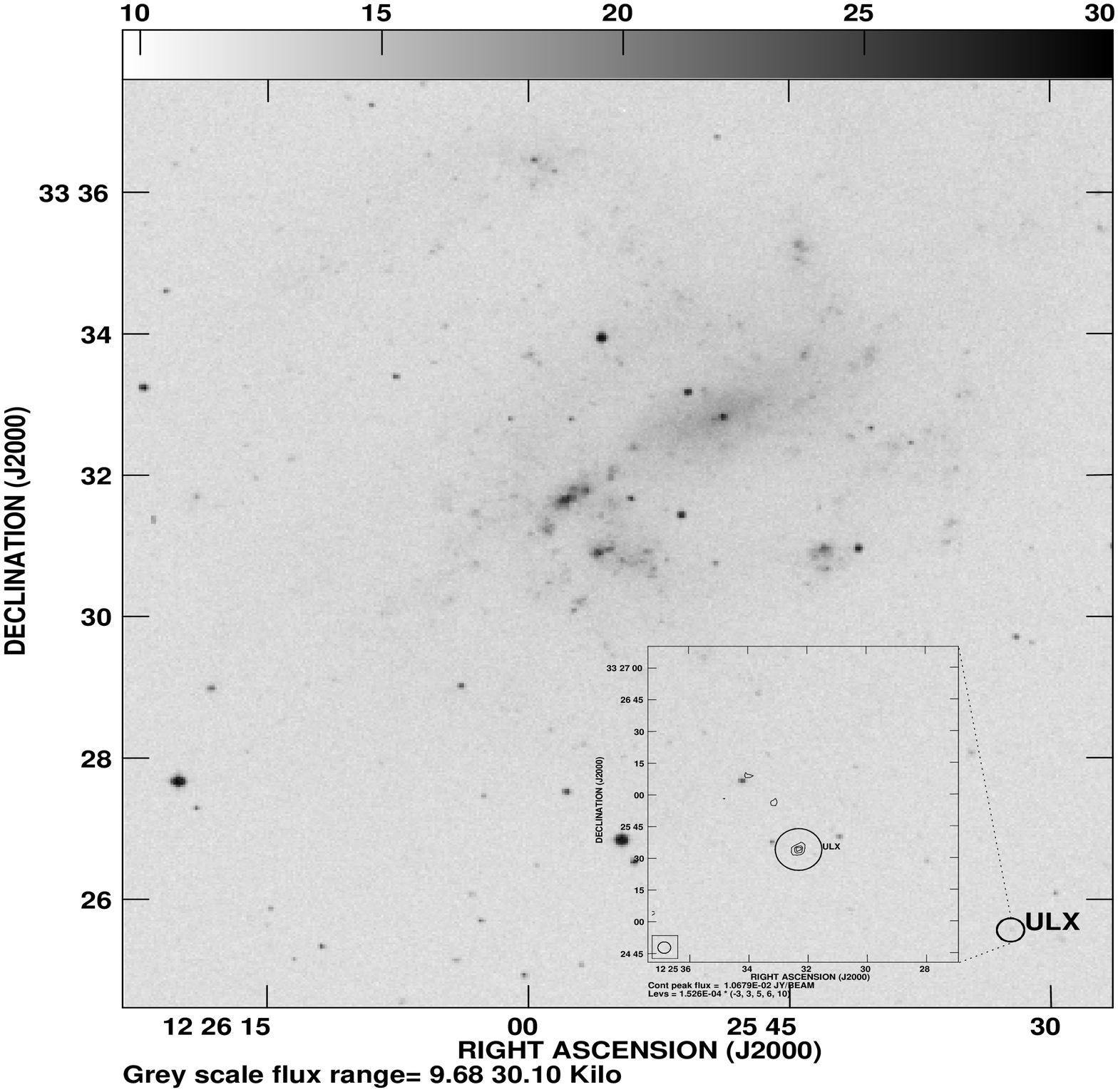}}
      \caption{Radio counterpart candidate of the X2 source in the
      field of NGC 4395.}
      \label{ngc4395}
   \end{figure*}

\begin{figure*}
   \centering
\resizebox{\hsize}{!}{\includegraphics[angle=0]{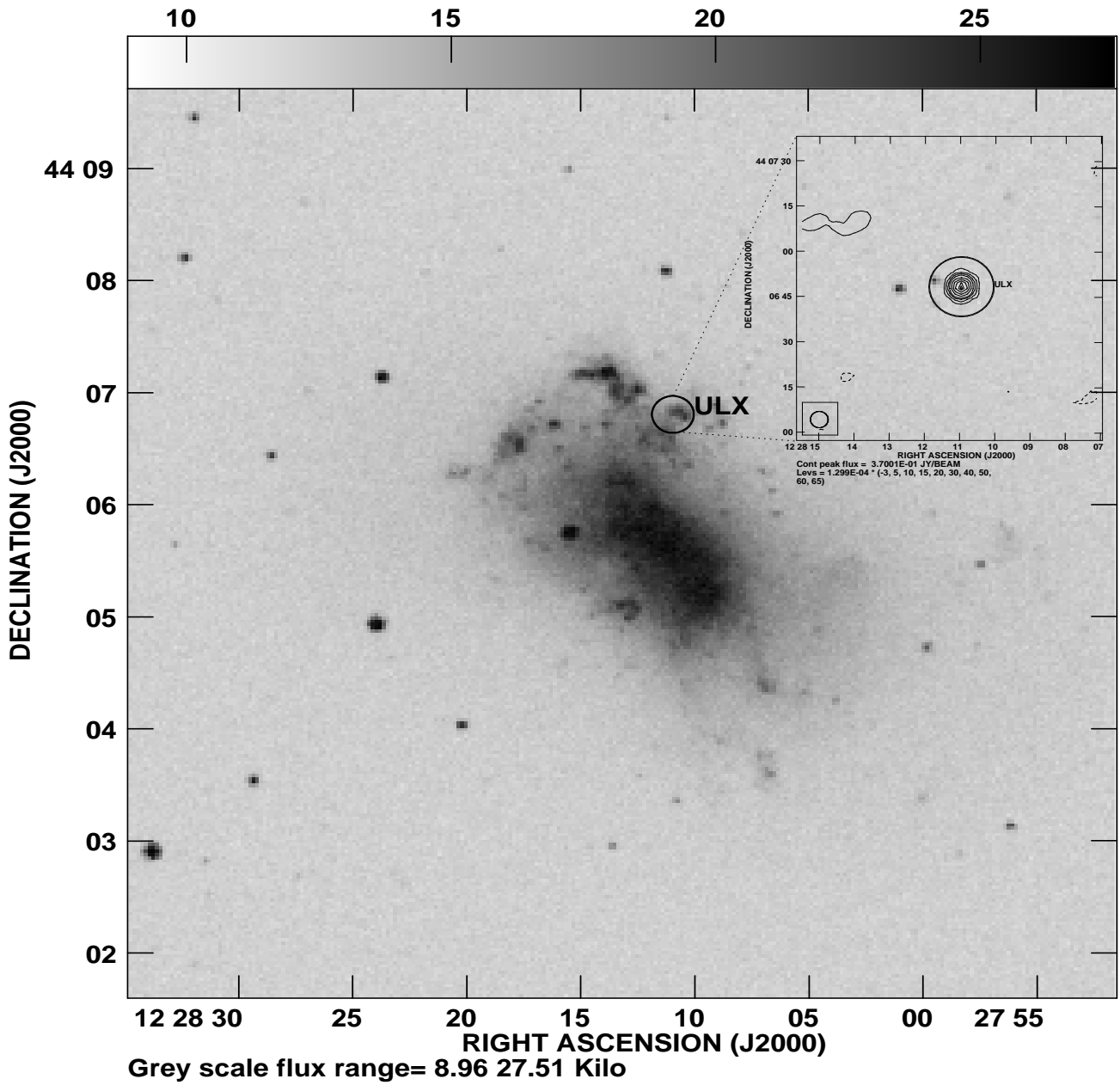}}
      \caption{Radio counterpart candidate of the X4 source in the
      field of NGC 4449.}
      \label{ngc4449}
   \end{figure*}

\begin{figure*}
   \centering
\resizebox{\hsize}{!}{\includegraphics[angle=0]{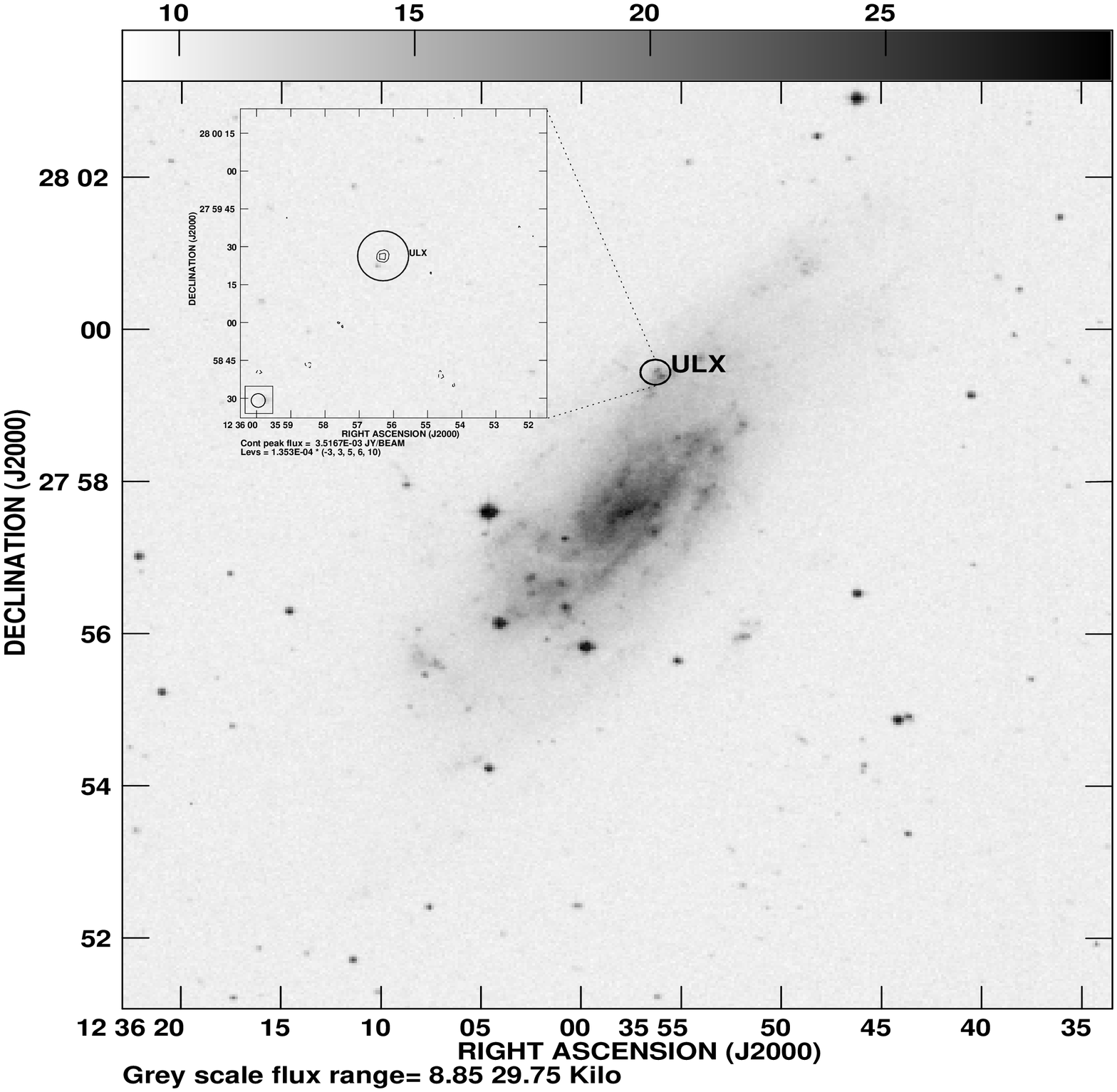}}
      \caption{Radio counterpart candidate of the X4 source in the
      field of NGC 4559.}
      \label{ngc4559}
   \end{figure*}

\begin{figure*}
   \centering
\resizebox{\hsize}{!}{\includegraphics[angle=0]{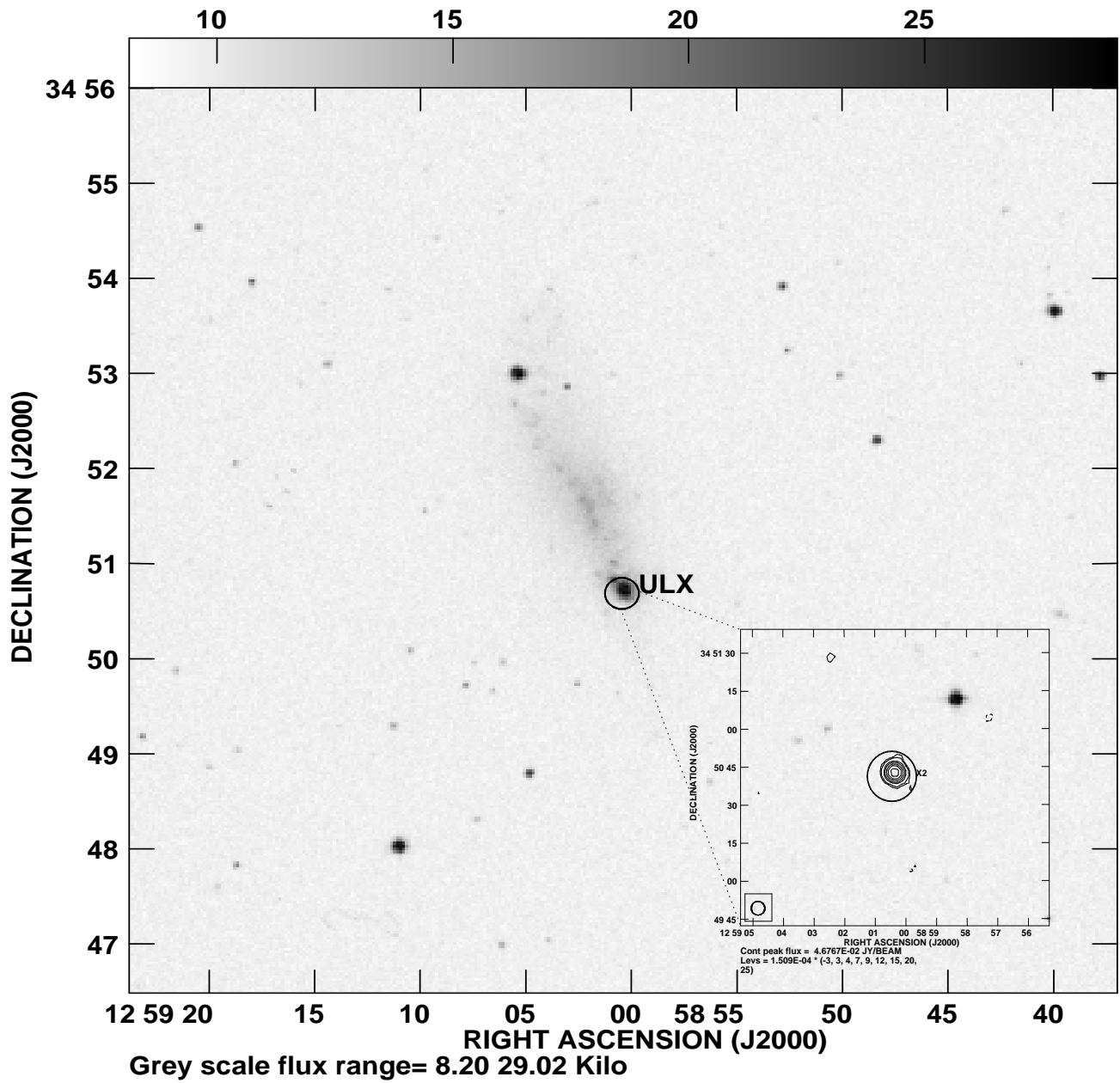}}
      \caption{Radio counterpart candidate of the X2 source in NGC
4861 in coincidence with a huge HII region.}
      \label{ngc4861}
   \end{figure*}

For each of these selected sources, we also show electronically in
Figs. \ref{ngc3738} to \ref{ngc5457} the optical and radio fields to
illustrate their positional coincidence. The optical images were
retrieved from the Digitized Sky Survey (DSS) (Lasker et al. 1990),
while the radio ones were taken from FIRST.  The following section is
devoted to providing a more detailed individual discussion.  It has
been necessary to adopt a distance for the different studied galaxies
as indicated in the electronic Table \ref{distances}. These values
were computed based on the Hubble law and assuming a Hubble constant
of $H_0=75$ km s$^{-1}$ Mpc$^{-1}$.

\begin{table*}
\caption[]{\label{coin} Selected non-nuclear, compact, and unidentified candidate radio counterparts of ULX sources}
\begin{tabular}{ccccccccc}
\hline
\hline
NGC    &  ULX  &  $\alpha_{{\rm J2000.0}}$  & $\delta_{{\rm J2000.0}}$ &  Radio vs.  &   Peak         &  Integrated  & Procedence &   Comments    \\
       &  Name &         (FIRST)            &        (FIRST)           &  X-ray      &  Flux Density  &  Flux Density    & catalogue(s) &            \\
       &       &                            &                          &  offset     &    (mJy)       &     (mJy)        &            &            \\
\hline
3738  & X1    & $11^h 35^m$45\rl 42 & $+54^{\circ} 33^{\prime}$15\pri 4 & 0\pri 06       &     61.37            &   94.11                  & LB         &  QSO?      \\
4088  & X1    & $12^h 05^m$31\rl 70 & $+50^{\circ} 32^{\prime}$46\pri 8 & 3\pri 62       &      1.87            &    5.07                  & LB, LM, CP &  HII region?         \\
4258  & X6    & $12^h 18^m$46\rl 32 & $+47^{\circ} 14^{\prime}$20\pri 4 & 3\pri 90       &      7.14            &   12.44                  & LB         &  QSO?                \\
4395  & X2    & $12^h 25^m$32\rl 31 & $+33^{\circ} 25^{\prime}$34\pri 0 & 0\pri 05       &      1.35            &    1.28                  & LB         &  $-$                \\
4449  & X4    & $12^h 28^m$10\rl 96 & $+44^{\circ} 06^{\prime}$48\pri 4 & 3\pri 30       &      9.27            &    9.81                  & LB         &  SNR?              \\
4559  & X4    & $12^h 35^m$56\rl 30 & $+27^{\circ} 59^{\prime}$26\pri 4 & 5\pri 35       &      1.12            &    0.90                  & LB         &  HII region?       \\
4861  & X2    & $12^h 59^m$00\rl 35 & $+34^{\circ} 50^{\prime}$42\pri 9 & 1\pri 97       &      3.88            &    6.01                  & LB, LM, CP &  HII region        \\
5457  & X9    & $14^h 03^m$41\rl 42 & $+54^{\circ} 19^{\prime}$05\pri 2 & 3\pri 40       &      6.63            &   11.45                  & LB, LM     &  HII region?       \\
5457  & X17   & $14^h 02^m$28\rl 12 & $+54^{\circ} 16^{\prime}$27\pri 3 & 4\pri 30       &      1.10            &    1.97                  & LB, LM     &  HII region?       \\
5457  & X26   & $14^h 04^m$29\rl 14 & $+54^{\circ} 23^{\prime}$53\pri 4 & 4\pri 07       &      2.65            &    4.05                  & LB, LM     &  HNR? SSS?         \\
5457  & X29   & $14^h 04^m$00\rl 78 & $+54^{\circ} 09^{\prime}$11\pri 3 & 6\pri 36       &      3.68            &    5.43                  & LB         &  FRII?             \\
\hline
\end{tabular}
\end{table*}

\section{Comments on selected non-nuclear coincidences}

X1 in NGC 3738.

The Sd irregular galaxy NGC 3738 is a member of a non interacting
pair.  The FIRST survey shows a double radio source at nearly two
arc-minutes from the galaxy nucleus. Only one radio component is
consistent with the ROSAT HRI position of the ULX source (see
electronic Fig. \ref{ngc3738}).  As mentioned above, ULX radio
counterparts are likely to be compact objects in the FIRST images.
Therefore, the detection of a double radio source strongly suggests
that X1 in NGC 3738 is not a ULX in this galaxy but is an unrelated
background object, possibly a QSO. This suggestion is further
supported by both its strong radio flux density, which outshines the
galaxy itself, and the ULX's very peripheral location.

X1 in NGC 4088.

The Holmberg SBc galaxy NGC 4088 forms a pair with NGC 4085 with both
belonging to the Ursa Major Cluster.  The ULX source is within the
extended emission of a spiral arm in NGC 4088 as shown in electronic
Fig. \ref{ngc4088}. Nevertheless, the X-ray position is coincident
with a conspicuous maximum of radio emission.  In spite of this fact
and based on the available data, we cannot rule out that most of the
radio emission comes from an HII region in the spiral arm.

X6 in NGC 4258.  

The nearby and almost edge-on barred (SAB(s)bc) galaxy NGC 4258
(Messier 106) hosts a weak AGN. It also forms a pair with NGC 4248 in
the Ursa Major Cluster.  The optical image (see electronic
Fig. \ref{ngc4258}) shows an apparent chain of HII regions at the
extremities of its arms. This galaxy is well known for its nuclear
thin disk in Keplerian rotation, which is traced by emission from
water masers (Miyoshi et al. 1995) and implies a supermassive central
black hole.

This case is reminiscent of the first object discussed above, both in
flux density level and galactocentric distance. The radio morphology
suggests a core+jet structure with the brighter component being
consistent with the X-ray position. Again, we speculate that this
object is most likely a background QSO unrelated to the galaxy.

X2 in NGC 4395.

The galaxy NGC 4395 has a loose and disconnected appearance containing
the faintest and nearest Seyfert 1 nucleus currently known and
classified as Sc/Irr type.  The proposed counterpart of X2 in NGC 4395
is a very compact and faint radio source in the peripheral outskirts
of the galaxy (electronic Fig. \ref{ngc4395}). With the present data
it is difficult to assess whether the object is associated to NGC 4395
itself.

X4 in NGC 4449.

The very blue Magellanic irregular galaxy NGC 4449 contains a lot of
HII regions.  In this case, the ULX radio counterpart candidate is a
strictly compact radio source projected against the galactic disk
(electronic Fig. \ref{ngc4449}).

In the FIRST image, it outshines the galactic nucleus. Thus, the
possibility that we are again dealing with a background object cannot
be ruled out. Alternatively, the ULX source could be associated with
some of the apparent extended regions seen at optical wavelengths
within the X-ray error box.  In fact, this ULX source corresponds to
source 15 in the Chandra results reported by \cite{sum2003}, who
assumed it to be a supernova remnant (SNR) based on the old VLA
observations of \cite{bs1983} with angular resolution poorer than
FIRST.

X4 in NGC 4559.

The Sc spiral galaxy NGC 4559 exhibits some easily resolved HII
regions.  X4 coincides with a compact and faint radio counterpart
candidate in the FIRST image as illustrated in electronic
Fig. \ref{ngc4559}.  It is clearly located within one of the spiral
arms. At optical wavelengths there is a possible HII region near the
ULX position that could be associated to. However, no extended radio
emission is detected around it.

X2 in NGC 4861.

The galaxy NGC 4861 is part of the Mrk 59 complex, together with the
dwarf irregular galaxy IC 3961.  In particular, NGC 4861 refers to the
bright knot in the southern region of the complex (see electronic
Fig. \ref{ngc4861}).  It has been suggested that it is an HII region
powered by massive early type stars.  The proposed radio counterpart
is a barely extended radio source very close to it.

X9, X17, X26, and X29 in NGC 5457.

Also known as Messier 101, NGC 5457 is a nearby face-on Sc spiral
galaxy within a rich group.  It is the prototype of the multiple-arm
galaxies of the Sc classification.

\begin{figure*}
   \centering
\resizebox{\hsize}{!}{\includegraphics[angle=0]{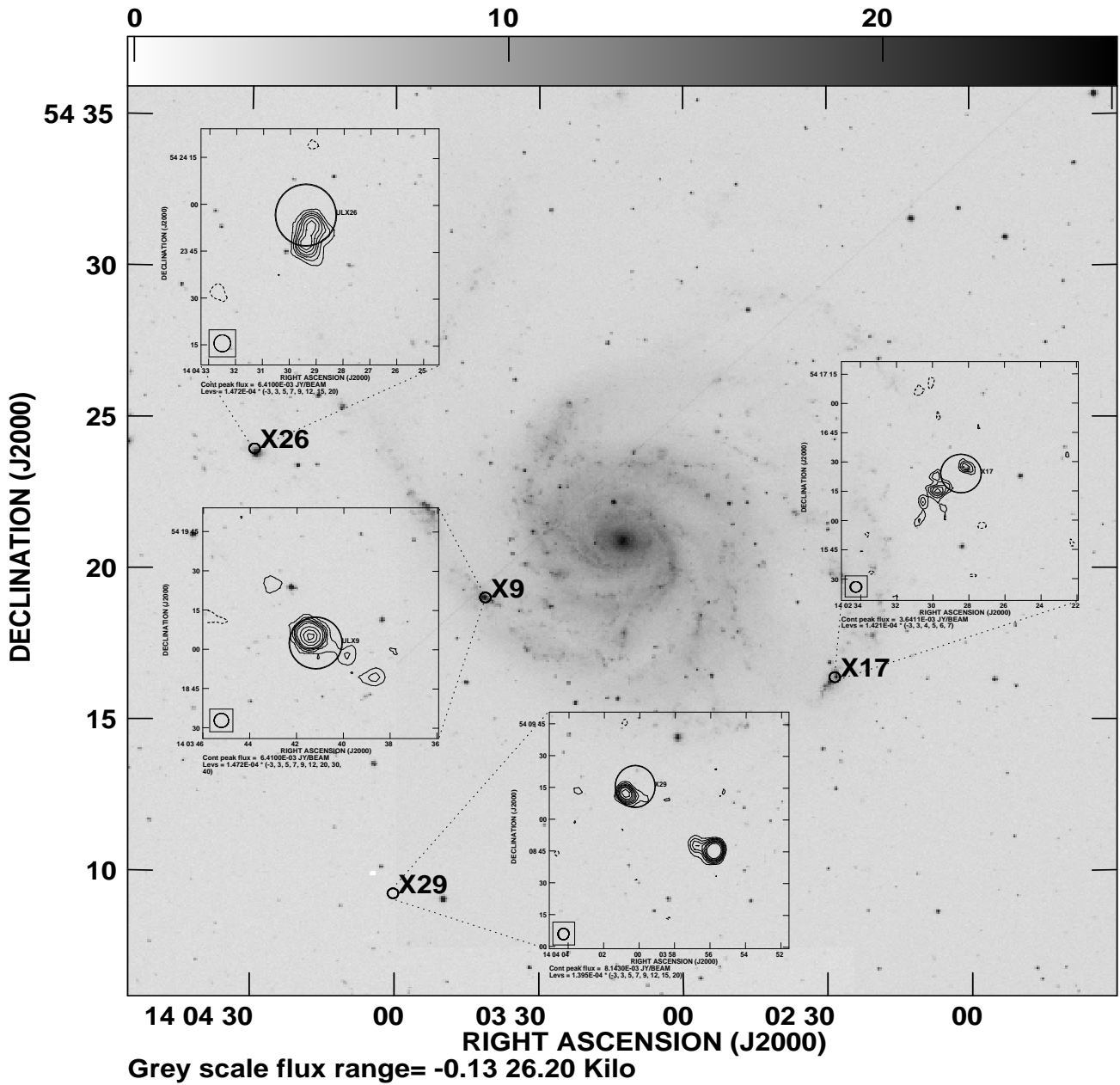}}
      \caption{Several ULX sources (X9, X17, X26, and X29) with new
FIRST radio counterpart candidates in the field of NGC 5457 (M~101).}
      \label{ngc5457}
   \end{figure*}

X9 and X17 appear projected against the spiral arms of the galactic
disk (see electronic Fig. \ref{ngc5457}).  Extended emission around
their proposed radio counterparts is clearly detected.  Therefore, the
association with these regions of star formation cannot be ruled out
in these cases.  In contrast, X26 and X29 in Messier 101 have radio
counterpart candidates with apparently double morphology (see again
Fig. \ref{ngc5457}).  They are located in the peripheral region of the
galaxy, so we could also be dealing with background sources as in the
first ULX case discussed in this section.  In particular, X29 reminds
of a Fanaroff-Riley Type II radio galaxy. On the other hand, X26 seems
to overlap with a fuzzy optical object (NGC 5471B) that has been
proposed by \cite{jen2004} to be either an hypernova remnant (HNR) or
a supersoft source (SSS) of X-ray binary type.

\section{Discussion}

We notice that 6 out of the 11 selected sources in Table \ref{coin}
appear to overlap with spiral arms or HII regions in their
corresponding galaxy. This fact is not new (Fabbiano 2004; Liu \&
Bregman 2001) and could be an indication of ULX sources forming
predominantly in star formation regions. Such a location would support
the idea of ULX sources being related to end products of fast evolving
massive stars. Thus, if IMBH exist, some of them could be linked to
the final stages of very high mass stars (Madau \& Rees 2001).  On the
other hand, it also cannot be ruled out that some of our objects are
merely background AGNs or QSOs and, therefore, they are related to
supermassive black holes. The number of wrong identifications in
\cite{lb2005} can be tentatively estimated by extrapolating our sample
over the entire catalogue.  Given that Table \ref{coin2} contains at
least three possible QSOs, this would imply a minimum of 24 QSOs
falsely identified as ULXs. More realistically, this number could
represent more than half of the total catalogue entries if radio-loud
nuclear coincidences were also considered as non-secure
identifications.

A possible way to try to constrain the nature of sources in Table
\ref{coin} is to study how they behave with respect to the correlation
between radio ($L_{\rm radio}$) and X-ray ($L_X$) luminosity and black
hole mass ($M$) proposed by several authors for black holes with
sub-Eddington accretion rates.  Such a correlation appears to be valid
for stellar, supermassive, and intermediate mass black holes, and both
observational and theoretical evidence seems to support it (see
e.g. K\"ording et al.~2002; Corbel et al.~2003; Gallo et al.~2003;
Merloni et al.~2003; Falcke et al.~2004 and references therein).  In
these systems, the spectral energy distribution is dominated by the
non-thermal emission from a relativistic jet.  The spectrum can be
approximated by a broken power law with a flat-to-inverted optically
thick part, typically in the radio domain, and with a steep optically
thin part at X-ray energies.  In this context ULXs would be low-mass
analogs of BL Lac objects.  Using the \cite{fal2004} formulation, the
non-linear correlation can be generally expressed as
\begin{equation}
L_X \propto L_R^m M^{\alpha_X - m \alpha_R} \label{correleq}
\end{equation} 
where $m=(17/12-2\alpha_X /3)/(17/12 - 2\alpha_R/3)$ and, on average,
the radio optically thick and X-ray optically thin spectral indexes
are $\alpha_R \simeq 0.15$ and $\alpha_X \simeq -0.6$,
respectively. For practical purposes we adopt these values so
Eq. \ref{correleq} then becomes
\begin{equation}
L_X \propto L_R^{1.38} M^{-0.81}.
\end{equation}

\begin{figure*}
   \centering
\resizebox{\hsize}{!}{\includegraphics[angle=0]{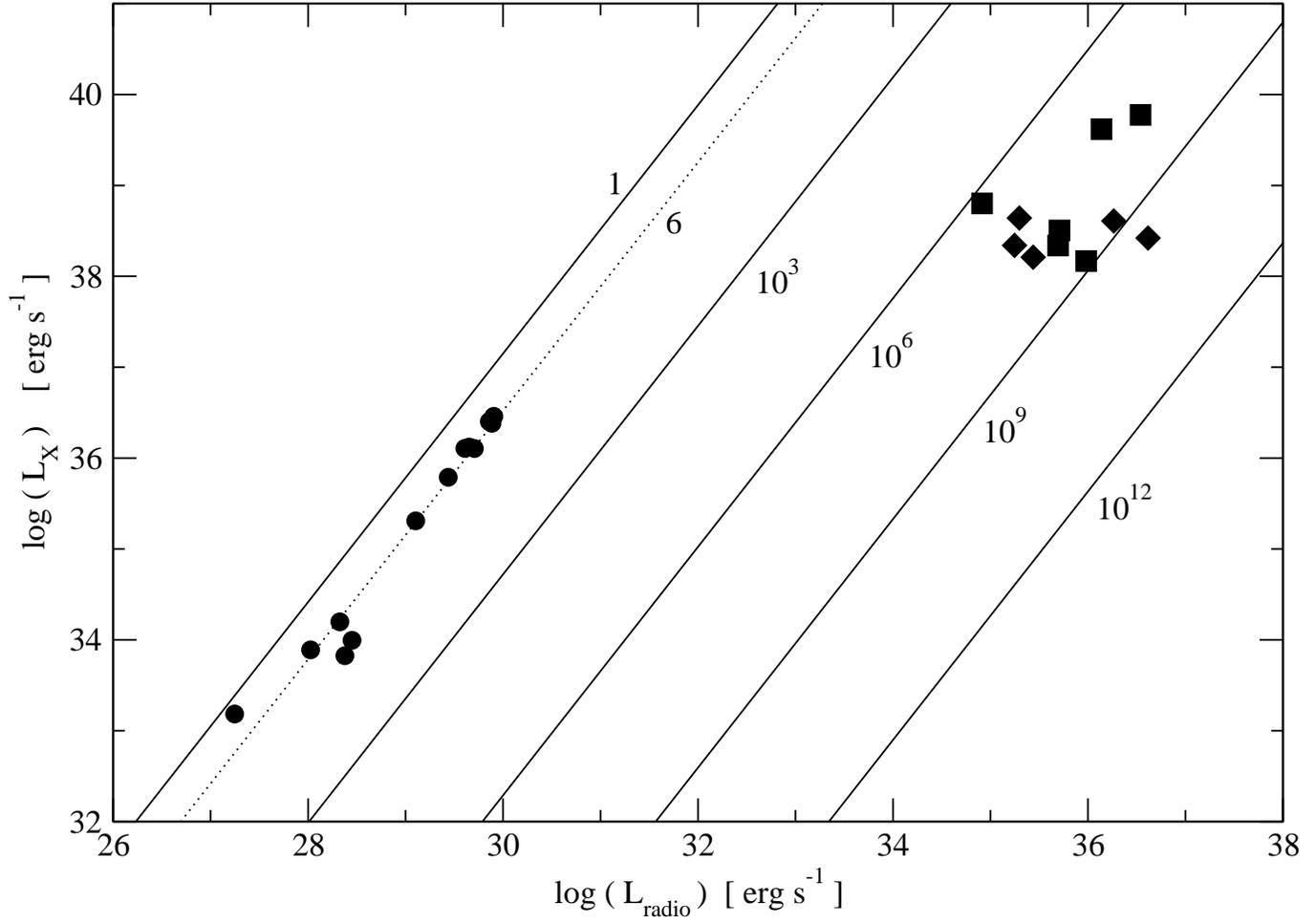}}
      \caption{Comparison of the ULX sources studied with the
correlation in the $L_X$--$L_{\rm radio}$ plane assuming a
sub-Eddington radio-loud black-hole scenario.  The parallel lines
correspond to the labeled black hole mass relative to that of the Sun.
The filled circles are the \cite{cor2003} data for the 6 $M_{\odot}$
black hole X-ray binary GX339$-$4 where such a correlation is very
well established. Diamonds and squares represent peripheral and
non-peripheral ULX sources, respectively.  }
      \label{correl}
   \end{figure*}

To test the correlation for our ULX sources we have used the distances
from electronic Table \ref{distances} as quoted before. The X-ray
luminosities were taken from \cite{lb2005}, while the monochromatic
radio luminosities are those derived from FIRST flux densities at 1.4
GHz. The \cite{fal2004} correlation parameters were established using
the X-ray energy band of 2-10 keV and the monochromatic radio
luminosity at 5 GHz. For consistency, we have consequently scaled our
$L_X$ and $L_{\rm radio}$ values by using the spectral index values
just quoted above, and the resulting plot is shown in electronic
Fig. \ref{correl}.

It can be seen here that none of the selected ULX sources studied
would apparently correspond to an IMBH.  In the framework of
correlations in the sub-Eddington accretion regime, this suggests that
most of these objects would be background AGNs unrelated to their
respective galaxy, as we previously thought.  Independent of the
correlation, such a statement is especially likely in cases of double
radio sources in NGC 3738, NGC 4258, and NGC 5457.  A true ULX nature
for some of the other sources is not strictly excluded especially if
other physical scenarios are considered, such as strongly beamed
stellar sources. Further observations with higher angular resolution
and sensitivity will be certainly needed before a new radio emitting
ULX source of the one reported in this work could confidently be
claimed to be a true ULX.

\section{Conclusions}

We have presented a catalogue of 70 ULX coincidences with FIRST radio
sources based on a cross-identification.  This constitutes an example
of how modern available databases and archives are powerful resources
still to be exploited towards solving different issues in today's
astrophysics. Most entries are likely nuclear coincidences with their
respective galaxy and, therefore, not true ULX in a strict sense. We
do find 11 cases of non-nuclear ULXs with new radio counterpart
candidates that have got our attention.

Among them, however, most are probably background QSOs or AGNs and not
proper ULX sources.  Double radio morphology with typical quasar or
radio galaxy appearance supports this statement in at least 4
objects. For the rest, the background hypothesis is further suggested
based on their behaviour with respect to the correlation between X-ray
and radio luminosities and the mass of the black hole accreting at
sub-Eddington regimes. In particular, our correlation plot indicates
supermassive black holes ($\sim 10^6$-$10^{9}$ $M_{\odot}$), and these
values are not expected at all in a galaxy disk, given the peripheral
location of the sources.

Nevertheless, a true ULX nature is actually not excluded, at least in
cases without double radio-counterpart candidates.  In fact, nearly
half of these candidates appear compact or extended and are projected
against HII regions in the galaxy spiral arms.  This could be
consistent with physical scenarios where ULX sources are linked to end
products of very massive star evolution or binary systems with strong
beaming effects. Further investigation will be needed to shed more
light on this open question.

\begin{acknowledgements}
The authors acknowledge support by the DGI of the Ministerio de
Educaci\'on y Ciencia (Spain) under grant AYA2004-07171-C02-02, FEDER
funds, and Plan Andaluz de Investigaci\'on of Junta de Andaluc\'{\i}a
as research group FQM322.  This research made use of the SIMBAD
database, operated at the CDS, Strasbourg, France.  The research of
DPR has been supported by the Education Council of Junta de
Andaluc\'{\i}a (Spain).  We also thank Josep M. Paredes (Univ. of
Barcelona) for his helpful comments and discussion.
\end{acknowledgements}

\Online

\begin{table*}
\caption[]{\label{coin2} Candidate radio counterparts of ULX sources found after the cross-identification process}
\tiny
\begin{tabular}{cccccccccc}
\hline
\hline
NGC    &  ULX  &  $\alpha_{{\rm J2000.0}}$  & $\delta_{{\rm J2000.0}}$ &  Radio vs.  &   Peak         &  Integrated      &  X-ray         & Procedence &   Comments$^{1}$    \\
       &  Name &         (FIRST)            &        (FIRST)           &  X-ray      &  Flux Density  &  Flux Density    &  Luminosity    & catalog(s) &            \\
       &       &                            &                          &  offset     &    (mJy)       &     (mJy)        & (10$^{39}$ erg s$^{-1}$)  &            &            \\
\hline
1052  & X1    & $02^h 41^m$04\rl 81 & $-08^{\circ} 15^{\prime}$20\pri 9 & 2\pri 65       &    967.08            & 1017.22              &    32.8               & LB         & N      \\
1068  & X1    & $02^h 42^m$40\rl 65 & $-00^{\circ} 00^{\prime}$49\pri 2 & 5\pri 25       &    797.54            & 1917.93              &  1330                 & LB, LM     & N      \\
1073  & X1    & $02^h 43^m$39\rl 60 & $+01^{\circ} 21^{\prime}$09\pri 5 & 2\pri 92       &     39.25            &   55.98              &     3.16              & LB         & QSO    \\
2782  & X1    & $09^h 14^m$05\rl 08 & $+40^{\circ} 06^{\prime}$49\pri 4 & 2\pri 18       &     34.48            &  107.44              &    98.5               & LB         & N      \\
2903  & X1    & $09^h 32^m$10\rl 09 & $+21^{\circ} 30^{\prime}$04\pri 3 & 2\pri 24       &     19.03            &   63.39              &     5.26              & LB         & N      \\
2974  & X1    & $09^h 42^m$33\rl 28 & $-03^{\circ} 41^{\prime}$56\pri 9 & 0\pri 24       &      5.88            &    7.54              &    31.7               & LB         & N      \\
3190  & X1    & $10^h 18^m$05\rl 63 & $+21^{\circ} 49^{\prime}$55\pri 1 & 6\pri 65       &      2.50            &    4.03              &     8.79              & LB         & N      \\
3226  & X1    & $10^h 23^m$27\rl 01 & $+19^{\circ} 53^{\prime}$54\pri 8 & 1\pri 96       &      4.45            &    4.49              &    18                 & LB         & N      \\
3227  & X1    & $10^h 23^m$30\rl 58 & $+19^{\circ} 51^{\prime}$54\pri 5 & 1\pri 12       &     74.39            &   82.85              &   489                 & LB         & N      \\
3310  & X1    & $10^h 38^m$45\rl 97 & $+53^{\circ} 30^{\prime}$14\pri 3 & 4\pri 38       &      7.68            &  102.87              &    83.1               & LB, LM     & N      \\
3395  & X1    & $10^h 49^m$50\rl 14 & $+32^{\circ} 59^{\prime}$00\pri 1 & 7\pri 18       &      1.28            &   20.64              &     5.62              & LM         & N      \\
3627  & X1    & $11^h 20^m$15\rl 00 & $+12^{\circ} 59^{\prime}$29\pri 6 & 8\pri 31       &     12.01            &   13.40              &     2.21              & LB         & N      \\
3665  & X1    & $11^h 24^m$43\rl 03 & $+38^{\circ} 45^{\prime}$51\pri 9 & 9\pri 87       &     12.47            &   24.10              &    33.5               & LB         & N      \\
3690  & X2    & $11^h 28^m$30\rl 95 & $+58^{\circ} 33^{\prime}$40\pri 8 & 4\pri 16       &     56.22            &   83.49              &     6.6               & LM         & N$^{2}$ \\
3690  & X3    & $11^h 28^m$30\rl 95 & $+58^{\circ} 33^{\prime}$40\pri 8 & 4\pri 20       &     56.22            &   83.49              &    23.9               & LM         & N$^{2}$ \\
3690  & X8    & $11^h 28^m$33\rl 64 & $+58^{\circ} 33^{\prime}$46\pri 2 & 4\pri 13       &    169.31            &  282.27              &     3.5               & LM         & N$^{2}$ \\
3690  & X10   & $11^h 28^m$33\rl 64 & $+58^{\circ} 33^{\prime}$46\pri 2 & 2\pri 04       &    169.31            &  282.27              &     5.8               & LM         & N$^{2}$ \\
3690  & X11   & $11^h 28^m$33\rl 64 & $+58^{\circ} 33^{\prime}$46\pri 2 & 5\pri 07       &    169.31            &  282.27              &     4.1               & LM         & N$^{2}$ \\
3690  & X12   & $11^h 28^m$33\rl 64 & $+58^{\circ} 33^{\prime}$46\pri 2 & 7\pri 11       &    169.31            &  282.27              &     2.9               & LM         & N$^{2}$ \\
3738  & X1    & $11^h 35^m$45\rl 42 & $+54^{\circ} 33^{\prime}$15\pri 4 & 0\pri 06       &     61.37            &   94.11              &     0.371             & LB         & QSO?   \\
3982  & X1    & $11^h 56^m$28\rl 15 & $+55^{\circ} 07^{\prime}$31\pri 0 & 1\pri 34       &      3.07            &    3.97              &     5.13              & LB         & N      \\
3998  & X1    & $11^h 57^m$56\rl 13 & $+55^{\circ} 27^{\prime}$13\pri 0 & 1\pri 85       &     93.45            &   98.51              &   259                 & LB         & N      \\
4051  & X1    & $12^h 03^m$09\rl 59 & $+44^{\circ} 31^{\prime}$52\pri 5 & 2\pri 29       &     12.30            &   19.34              &  2770                 & LB         & N      \\
4088  & X1    & $12^h 05^m$31\rl 70 & $+50^{\circ} 32^{\prime}$46\pri 8 & 3\pri 62       &      1.87            &    5.07              &     5.86              & LB, LM, CP & HII?   \\          
4151  & X1    & $12^h 10^m$32\rl 55 & $+39^{\circ} 24^{\prime}$21\pri 0 & 0\pri 24       &    290.78            &  331.29              &  1330                 & LB         & N      \\
4156  & X1    & $12^h 10^m$49\rl 61 & $+39^{\circ} 28^{\prime}$22\pri 2 & 1\pri 28       &      2.46            &    2.93              &     4.32              & LB         & N      \\
4203  & X1    & $12^h 15^m$05\rl 06 & $+33^{\circ} 11^{\prime}$50\pri 7 & 1\pri 25       &      6.94            &    7.72              &   109                 & LB         & N      \\
4235  & X1    & $12^h 17^m$09\rl 86 & $+07^{\circ} 11^{\prime}$29\pri 7 & 3\pri 00       &      4.66            &    5.10              &  1270                 & LB         & N      \\
4258  & X6    & $12^h 18^m$46\rl 32 & $+47^{\circ} 14^{\prime}$20\pri 4 & 3\pri 90       &      7.14            &   12.44              &     0.572             & LB         & QSO?   \\
4261  & X1    & $12^h 19^m$23\rl 23 & $+05^{\circ} 49^{\prime}$29\pri 9 & 3\pri 99       &    135.57            &  140.46              &   229                 & LB         & N      \\
4278  & X1    & $12^h 20^m$06\rl 82 & $+29^{\circ} 16^{\prime}$50\pri 7 & 1\pri 97       &    389.38            &  402.00              &    42.8               & LB         & N      \\
4303  & X1    & $12^h 21^m$54\rl 91 & $+04^{\circ} 28^{\prime}$26\pri 7 & 1\pri 80       &      4.85            &    9.74              &     8.51              & LB         & N      \\
4321  & X1    & $12^h 22^m$54\rl 95 & $+15^{\circ} 49^{\prime}$20\pri 7 & 5\pri 17       &      3.40            &   37.35              &    17.2               & LB, LM     & N      \\
4321  & X2    & $12^h 22^m$58\rl 66 & $+15^{\circ} 47^{\prime}$51\pri 8 & 2\pri 50       &      5.18            &    5.64              &     1.57              & LB         & SNR    \\
4374  & X1    & $12^h 25^m$03\rl 72 & $+12^{\circ} 53^{\prime}$21\pri 8 & 7\pri 77       &    105.14            &  196.92              &   137                 & LB         & N      \\
4388  & X1    & $12^h 25^m$46\rl 74 & $+12^{\circ} 39^{\prime}$42\pri 7 & 7\pri 56       &     24.35            &   35.84              &    23                 & LB         & N      \\
4395  & X2    & $12^h 25^m$32\rl 31 & $+33^{\circ} 25^{\prime}$34\pri 0 & 0\pri 05       &      1.35            &    1.28              &     0.308             & LB         & $-$    \\
4435  & X1    & $12^h 27^m$40\rl 54 & $+13^{\circ} 04^{\prime}$44\pri 9 & 3\pri 45       &      1.73            &    2.16              &     6.67              & LB         & N      \\
4438  & X1    & $12^h 27^m$45\rl 57 & $+13^{\circ} 00^{\prime}$32\pri 9 & 2\pri 43       &     56.18            &   76.24              &    25.7               & LB         & N      \\
4449  & X4    & $12^h 28^m$10\rl 96 & $+44^{\circ} 06^{\prime}$48\pri 4 & 3\pri 30       &      9.27            &    9.81              &     0.45              & LB         & S?     \\
4472  & X1    & $12^h 29^m$46\rl 74 & $+08^{\circ} 00^{\prime}$01\pri 6 & 5\pri 25       &     98.39            &  123.70              &   181                 & LB         & N      \\
4485  & X3    & $12^h 30^m$31\rl 40 & $+41^{\circ} 39^{\prime}$08\pri 6 & 7\pri 16       &      1.29            &   12.88              &     4.6               & LM         & E      \\
4526  & X1    & $12^h 34^m$02\rl 93 & $+07^{\circ} 41^{\prime}$58\pri 6 & 7\pri 13       &      3.76            &   12.00              &     7.16              & LB         & N      \\
4536  & X1    & $12^h 34^m$27\rl 07 & $+02^{\circ} 11^{\prime}$18\pri 0 & 2\pri 44       &     44.80            &  131.82              &     5.76              & LB         & N      \\
4552  & X1    & $12^h 35^m$39\rl 80 & $+12^{\circ} 33^{\prime}$22\pri 9 & 3\pri 09       &    107.95            &  112.78              &    82.6               & LB         & N      \\
4559  & X4    & $12^h 35^m$56\rl 30 & $+27^{\circ} 59^{\prime}$26\pri 4 & 5\pri 35       &      1.12            &    0.90              &     0.207             & LB         & HII?   \\
4565  & X1    & $12^h 36^m$20\rl 77 & $+25^{\circ} 59^{\prime}$15\pri 8 & 1\pri 60       &      1.48            &    1.88              &    12.1               & LB         & N      \\
4569  & X1    & $12^h 36^m$49\rl 82 & $+13^{\circ} 09^{\prime}$46\pri 7 & 2\pri 13       &      7.69            &   14.36              &    13.3               & LB         & N      \\
4593  & X1    & $12^h 39^m$39\rl 44 & $-05^{\circ} 20^{\prime}$39\pri 2 & 4\pri 51       &      2.82            &    5.20              & 15600                 & LB         & N      \\
4631  & X2    & $12^h 42^m$10\rl 59 & $+32^{\circ} 32^{\prime}$37\pri 0 & 6\pri 57       &      3.93            &   38.34              &     1.34              & LM         & N$^{3}$ \\
4636  & X1    & $12^h 42^m$49\rl 96 & $+02^{\circ} 41^{\prime}$16\pri 2 & 4\pri 66       &     14.03            &   56.91              &   148                 & LB         & N      \\
4649  & X1    & $12^h 43^m$39\rl 98 & $+11^{\circ} 33^{\prime}$09\pri 6 & 3\pri 59       &     15.26            &   16.54              &   229                 & LB         & N      \\
4736  & X1    & $12^h 50^m$53\rl 06 & $+41^{\circ} 07^{\prime}$13\pri 6 & 3\pri 58       &      8.91            &   13.51              &    13.3               & LB         & N      \\
4826  & X1    & $12^h 56^m$43\rl 62 & $+21^{\circ} 41^{\prime}$00\pri 3 & 4\pri 13       &      7.95            &   36.06              &     4.49              & LB         & N      \\
4861  & X2    & $12^h 59^m$00\rl 35 & $+34^{\circ} 50^{\prime}$42\pri 9 & 1\pri 97       &      3.88            &    6.01              &     8.4               & LB, LM, CP & HII    \\
5005  & X1    & $13^h 10^m$56\rl 23 & $+37^{\circ} 03^{\prime}$33\pri 1 & 7\pri 97       &     35.90            &   45.81              &    58.3               & LB         & N      \\
5033  & X1    & $13^h 13^m$27\rl 50 & $+36^{\circ} 35^{\prime}$38\pri 1 & 1\pri 79       &      8.74            &   23.59              &   135                 & LB         & N      \\
5055  & X1    & $13^h 15^m$49\rl 21 & $+42^{\circ} 01^{\prime}$46\pri 3 & 1\pri 31       &      1.81            &   20.23              &     2.2               & LB         & N      \\
5194  & X1    & $13^h 29^m$52\rl 02 & $+47^{\circ} 11^{\prime}$43\pri 5 & 6\pri 62       &      1.26            &   16.54              &    17.2               & LB         & N      \\
5194  & X3    & $13^h 29^m$50\rl 65 & $+47^{\circ} 11^{\prime}$54\pri 6 & 0\pri 51       &      1.20            &    3.20              &     3.95              & LM         & N$^{3}$\\
5195  & X1    & $13^h 29^m$59\rl 52 & $+47^{\circ} 15^{\prime}$58\pri 5 & 4\pri 33       &      5.87            &    9.17              &     3.27              & LB         & N      \\
5273  & X1    & $13^h 42^m$08\rl 35 & $+35^{\circ} 39^{\prime}$15\pri 4 & 2\pri 29       &      2.60            &    2.90              &     4.72              & LB         & N      \\
5322  & X1    & $13^h 49^m$15\rl 12 & $+60^{\circ} 11^{\prime}$32\pri 8 & 8\pri 82       &      7.19            &    8.78              &    29.7               & LB         & N      \\
5350  & X1    & $13^h 53^m$21\rl 60 & $+40^{\circ} 21^{\prime}$50\pri 5 & 4\pri 27       &      2.85            &    4.03              &    21.4               & LB         & N      \\
5353  & X1    & $13^h 53^m$26\rl 69 & $+40^{\circ} 16^{\prime}$58\pri 9 & 6\pri 04       &     36.19            &   38.30              &   106                 & LB         & N      \\
5457  & X9    & $14^h 03^m$41\rl 42 & $+54^{\circ} 19^{\prime}$05\pri 2 & 3\pri 40       &      6.63            &   11.45              &     0.307             & LB, LM     & HII?   \\
5457  & X17   & $14^h 02^m$28\rl 12 & $+54^{\circ} 16^{\prime}$27\pri 3 & 4\pri 30       &      1.10            &    1.97              &     0.891             & LB, LM     & HII?   \\
5457  & X26   & $14^h 04^m$29\rl 14 & $+54^{\circ} 23^{\prime}$53\pri 4 & 4\pri 07       &      2.65            &    4.05              &     0.615             & LB, LM     & HNR? SSS?  \\
5457  & X29   & $14^h 04^m$00\rl 78 & $+54^{\circ} 09^{\prime}$11\pri 3 & 6\pri 36       &      3.68            &    5.43              &     0.228             & LB         & FRII?  \\        
5506  & X1    & $14^h 13^m$14\rl 88 & $-03^{\circ} 12^{\prime}$27\pri 5 & 1\pri 22       &    309.97            &  331.36              &   455                 & LB         & N      \\
\hline
\hline
\end{tabular}
~\\
(1) N: nuclear; QSO: quasar; HII: HII region; SNR: supernova remnant; E: non nuclear but extended; HNR: hypernova remnant; SSS: supersoft source; FRII: Fanaroff-Riley type II.   \\
(2) Multiple coincidences with a single FIRST source. \\
(3) Very close to nucleus, but not exactly coincident. \\

\end{table*}

\begin{table*}
\caption[]{\label{distances} Adopted distances to the galaxies listed in Table \ref{coin}}
\begin{tabular}{ccc}
\hline
\hline
NGC  &   Redshift ($\times 10^{-4}$) &     Distance  (Mpc)    \\
\hline
3738 &     7.64      &      3.05         \\
4088 &     25.24     &     10.09         \\
4258 &     14.94     &      5.97         \\
4395 &     10.64     &      4.25         \\
4449 &      6.9      &      2.76         \\
4559 &     27.22     &     10.88         \\
4861 &     27.79     &     11.11         \\
5457 &      8.04     &      3.21         \\
\hline
\hline
\end{tabular}
\end{table*}

\clearpage

\end{document}